\documentclass[reprint]{revtex4-1}

\usepackage{color}
\usepackage{graphicx}
\usepackage{epstopdf}
\usepackage{mathrsfs}
\usepackage{paralist}
\usepackage{enumerate}
\usepackage{amssymb,amsfonts,amsmath}
\graphicspath{{Figures/}}

\newcommand{\erm}{\mathrm{e}}
\newcommand{\irm}{\mathrm{i}}

\begin{document}


\title{Resonance-induced enhancement of the energy harvesting performance of piezoelectric flags} 



\author{Yifan \surname{Xia}}
\email[]{xia.yifan@ladhyx.polytechnique.fr}
\affiliation{LadHyX--D\'epartement de M\'ecanique, \'Ecole Polytechnique -- CNRS, Route de Saclay, 91128 Palaiseau, France}
\author{S\'ebastien \surname{Michelin}}
\affiliation{LadHyX--D\'epartement de M\'ecanique, \'Ecole Polytechnique -- CNRS, Route de Saclay, 91128 Palaiseau, France}
\author{Olivier \surname{Doar\'e}}
\affiliation{IMSIA, ENSTA ParisTech, CNRS, CEA, EDF, Universit\'e Paris-Saclay, 828 bd des Mar\'echaux, 91762 Palaiseau cedex France}

\date{\today}

\begin{abstract}
The spontaneous flapping of a flag can be used to produce electrical energy from a fluid flow when coupled to a generator. In this paper, the energy harvesting performance of a flag covered by a single pair of PVDF piezoelectric electrodes is studied both experimentally and numerically. The electrodes are connected to a resistive-inductive circuit that forms a resonant circuit with the piezoelectric's intrinsic capacitance. Compared with purely resistive circuits, the resonance between the circuit and the flag's flapping motion leads to a significant increase in the harvested energy. Our experimental study also validates our fluid-solid-electric nonlinear numerical model.
\end{abstract}


\maketitle 

Flow-induced instabilities and vibrations have recently received a renewed attention as potential mechanisms to produce electrical energy from ambient flows. Such instabilities enable a spontaneous and self-sustained motion of a solid body which can be used to  convert this mechanical energy into electrical form.\cite{bernitsas:2008, peng:2009,dias:2013}

The flapping of a flexible plate in an axial flow (thereafter referred to as a ``flag'') is a canonical example of such instabilities. Due to its rich and complex dynamics, this instability has been extensively studied during the last century.\cite{shelley:2011} The origin of this instability lies in a competition between the destabilizing fluid force and the stabilizing structural elasticity. The flag becomes unstable when the flow velocity exceeds a critical value, leading to a large amplitude self-sustained flapping.\cite{argentina:2005, eloy:2007, connell:2007, alben:2008c, michelin:2008, virot:2013}

Piezoelectric materials produce electric charge displacements when attached to a deformable structure,\cite{yang:2005} showing a ``direct piezoelectric effect'' that effectively qualifies them as electric generators. An output circuit connected to the electrodes of the piezoelectric elements can then exploit the generated electric current, as in vibration control applications.\cite{hagood:1991, thomas:2009} In the meantime, a feedback coupling is introduced by the inverse piezoelectric effect: any voltage between the electrodes creates an additional structural stress that would potentially influence the dynamics of the structure.

Piezoelectric energy generators received an increasing attention over the last two decades,\cite{williams:1996, umeda:1996} and a burgeoning research effort has been invested on this topic ever since.\cite{sodano:2004, anton:2007, erturkbook:2011, calio:2014piezoelectric} The work of Allen \& Smits using a piezoelectric membrane \cite{allen:2001} inspired several recent studies on flapping piezoelectric flags as energy-harvesting systems.\cite{dunnmon:2011, doare:2011, akcabay:2012, michelin:2013, xia:2015} Xia {\it et al.}\cite{xia:2015} investigated numerically the coupling between a piezoelectric flag and a resonant circuit, and identified an electro-mechanical frequency lock-in phenomenon where the output circuit dictates to the flag its flapping frequency and that significantly increases the energy-harvesting performance, compared with a simple resistive circuit.\cite{michelin:2013} This frequency lock-in phenomenon was however obtained for an idealized configuration, where (i) the flag is continuously covered with infinitesimally-small piezoelectric patches and (ii) the piezoelectric coupling is strong. The present work focuses on a configuration that is easier to obtain experimentally: the flag is covered by a single pair of piezoelectric patches made of Polyvinylidene Difluoride (PVDF), a material characterized by a relatively weak coupling. The energy-harvesting performance of this system is investigated both experimentally and numerically when connected to a resonant circuit.

\begin{figure}
\centering
\includegraphics{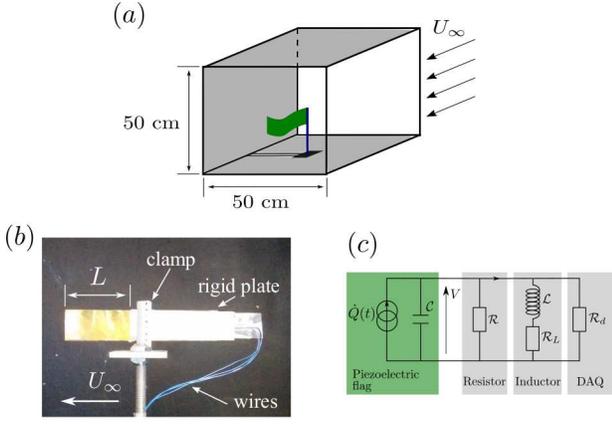}
\caption{\label{fig_flags_exp}$(a)$ PVDF piezoelectric flag placed in the wind tunnel with opaque walls (gray). $(c)$ Photo of the clamped PVDF flag. $(c)$ Equivalence of the harvesting circuit and data acquisition board (DAQ).}
\end{figure}

Experiments are performed using a PVDF piezoelectric flag. The PVDF film is cut into two patches of identical size: 9.5 cm $\times$ 2.5 cm. The two patches are then glued face-to-face with reverse polarity using a bonding tape, forming a flag covered by one single piezoelectric pair whose weight is 5.5$\times 10^{-4}$ kg. This flag is clamped at one end in a wind tunnel with a 50 cm $\times$ 50 cm test section (Fig.~\ref{fig_flags_exp}$a$). Using this single flag, the flag's ``effective length'', denoted by $L$, is varied by adjusting the position at which the flag is clamped. The remaining upstream section of the flag is attached to a rigid plate parallel to the flow so that its influence on the flow is minimized and the system effectively behaves as a flag of length $L$ (Fig.~\ref{fig_flags_exp}$b$). The electrodes of the flag are connected to a data acquisition board (DAQ) recording the voltage $V$, as well as an output circuit consisting of  a variable resistor $\mathcal{R}$ ranging from $5$ $\Omega$ to $10^8$ $\Omega$, and an inductor of inductance $\mathcal{L}$ in parallel connection.  From an electrical point of view, the piezoelectric flag is equivalent to a current source connected in parallel to an internal capacitance $\mathcal{C}$ and the equivalent circuit of the experimental setup is shown in Fig.~\ref{fig_flags_exp}$c$, where $\mathcal{R}_L$ and $\mathcal{R}_d$ are respectively the internal resistance of the inductor and the DAQ.
The mean harvested power ${P}$, defined as the energy dissipated in the output resistance $\mathcal{R}$ is obtained as:
\begin{equation}
\label{eq_Q_exp}
{P}=\langle V^2 \rangle\mathcal{R}^{-1},
\end{equation}
where $\langle\cdot \rangle$ is the time-averaging operator. The average is taken over 5 seconds, sufficiently long compared with the flag's flapping period (cf. Table~\ref{tab:two_lengths})

The numerical model presented in previous works\cite{michelin:2013, xia:2015} is adapted to the present configuration. Its main characteristics are briefly presented here and the reader is referred to Michelin \& Doar\'e\cite{michelin:2013} for more details. The flag of length $L$ and height $H$ is modeled as a clamped-free Euler-Bernoulli beam, whose mass per unit length is denoted by $\mu$, placed in a fluid of density $\rho$ flowing uniformly at velocity $U_\infty$. The fluid's forcing on the flag is computed using a local force model based on the relative motion between the flag and the flow.\cite{lighthill:1971, candelier:2011} Contrary to the previous studies referenced above, the flag is covered by one single pair of piezoelectric patches; consequently, the electric charge displacement $Q$ between the electrodes depends only on the flag's orientation at the trailing edge, $\Theta_T$, and on the voltage $V$, while the inverse piezoelectric effect introduces an added torque $\mathcal{M}_\text{piezo}$ applied locally at the flag's trailing edge. The relation between $Q$ and $V$ is prescribed by considering the electric circuit shown in Fig.~\ref{fig_flags_exp}$c$. 

Using $L$, $L/U_\infty$, $\rho HL^2$, $U_\infty\sqrt{\mu L/\mathcal{C}}$, $U_\infty\sqrt{\mu \mathcal{C} L}$ respectively as characteristic length, time, mass, voltage and electric charge, the system's electric state and the piezoelectric added torque $\mathcal{M}_\text{piezo}$ are given, in dimensionless forms by:
\begin{align}
\begin{split}
&\ddot{v}+\left( \beta_{L}\omega_0^2+\frac{1}{\beta_e} \right)\dot{v}+\left(1+\frac{\beta_{L}}{\beta_e}\right)\omega_0^2v\\
&\qquad\qquad\qquad\qquad
+\frac{\alpha}{U^*}\ddot{\theta_T}+\frac{\alpha}{U^*}\beta_{L}\omega_0^2\dot{\theta}_T=0,\label{eq:electric_state}
\end{split}\\
&\mathcal{M}_\text{piezo}=-\frac{\alpha}{U^*} v.\label{eq:inveffect_dim_sing}
\end{align}
Also, the problem is characterized by the following dimensionless parameters
\begin{equation}
\begin{aligned}
M^*&=\frac{\rho HL}{\mu},&U^*&=U_\infty\sqrt{\frac{\mu L^2}{B}},& H^*&=\frac{H}{L},\\
\alpha&=\chi\sqrt{\frac{L}{B\mathcal{C}}},&\beta&=\frac{\mathcal{R}U_\infty\mathcal{C}}{L},&\omega_0&=\frac{L}{U_\infty\sqrt{\mathcal{L}\mathcal{C}}},\label{ParaAd}
\end{aligned}
\end{equation}
with $M^*$ the fluid-solid inertia ratio, $U^*$ the reduced flow velocity, and $H^*$ the aspect ratio. The piezoelectric coupling coefficient, $\alpha$, characterizes the mutual forcing between the piezoelectric pair and the flag. Finally $\beta$ and $\omega_0$ characterize the resistive and inductive properties of the circuit. In the following, $\beta$, $\beta_d$, and $\beta_L$ correspond to the harvesting resistor $\mathcal{R}$, the DAQ's internal resistance $\mathcal{R}_d$, and the inductor's internal resistance $\mathcal{R}_L$, respectively. In Eq.~\eqref{eq:electric_state}, $\beta_e$ corresponds to the equivalent resistance to $\mathcal{R}$ and $\mathcal{R}_d$ connected in parallel:
\begin{equation}
\label{eq:beta_e}
\beta_e=\frac{\beta\beta_d}{\beta+\beta_d}.
\end{equation}

The coupling coefficient $\alpha$ defined in Eq.~\eqref{ParaAd} is determined from the conversion factor $\chi$, the bending rigidity $B$, and the capacitance $\mathcal{C}=15$ nF, which is measured using a multimeter. Note that because the whole PVDF flag (i.e. upstream and downstream of the clamp, Fig.~\ref{fig_flags_exp}$b$), is covered by two patches which are entirely connected in the circuit through their electrode, the total intrinsic capacitance $\mathcal{C}$ is independent of the effective length $L$. The bending rigidity $B$ of the flag is determined by measuring the flag's free vibration frequency $f_0$, which, at the first vibration mode, is given by: \cite{timoshenko1953history}
\begin{equation}
\label{eq:free_vib_freq}
f_0=\frac{3.515}{2\pi L^2}\sqrt{\frac{B}{\mu}}.
\end{equation}
The conversion factor $\chi$ is measured by connecting the flag uniquely to the DAQ, whose input impedance is $R_{d}=10^6$ $\Omega$. The voltage $V$ and deflection $\Theta_T$ are related through the direct piezoelectric effect,\cite{michelin:2013, xia:2015} which together with Ohm's law leads to:
\begin{equation}
\mathcal{R}_d\chi\dot{\Theta}_T+\mathcal{R}_d\mathcal{C}\dot{V}+V=0.
\end{equation}
Assuming purely harmonic signals, we write $V=V_0\erm^{ 2\irm\pi f t}$ and $\Theta_T=\Theta_0\erm^{\irm( 2\pi f t+\phi)}$, and using the previous equation:
\begin{equation}
\label{eq:chi_def}
\chi=\frac{V_0}{2\pi f \Theta_0}\sqrt{\frac{1}{R^2_{d}} +4\pi^2f^2 \mathcal{C}^2}.
\end{equation}

The measurements of the flapping frequency $f$, the amplitude of trailing edge orientation $\Theta_0$, and the amplitude of the voltage $V_0$ are performed in a second wind tunnel whose walls are transparent and allow for direct video recording of the flag's motion. This second wind tunnel is however not suitable for the rest of the present work due to its strong confinement (test section of 10 cm $\times$ 4 cm), which is not accounted for by the nonlinear numerical model used in this work. Confinement is however not a problem for the present measurements of $\chi$ since Eq.~\eqref{eq:chi_def} is valid regardless of the fluid forcing applied on the plate in the limit of linear piezoelectric coupling assumed here. Moreover, $\chi$ is an intrinsic property of the piezoelectric material that does not depend on the flow condition.

\begin{figure}
\centering
\includegraphics{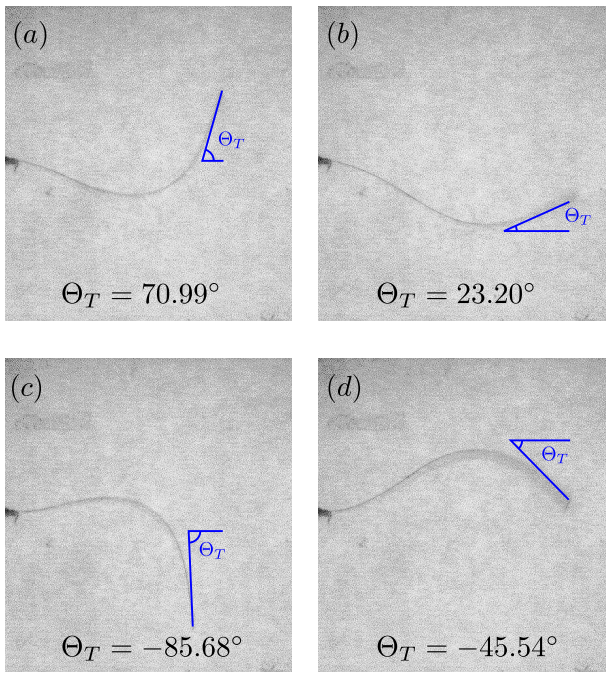}
\caption{\label{fig:measurement_Theta}Examples of measurement of $\Theta_T$ for $L=8$ cm and $U=12.3$ m/s in the transparent wind tunnel at four different instants.}
\end{figure}
\begin{figure}
\centering
\includegraphics{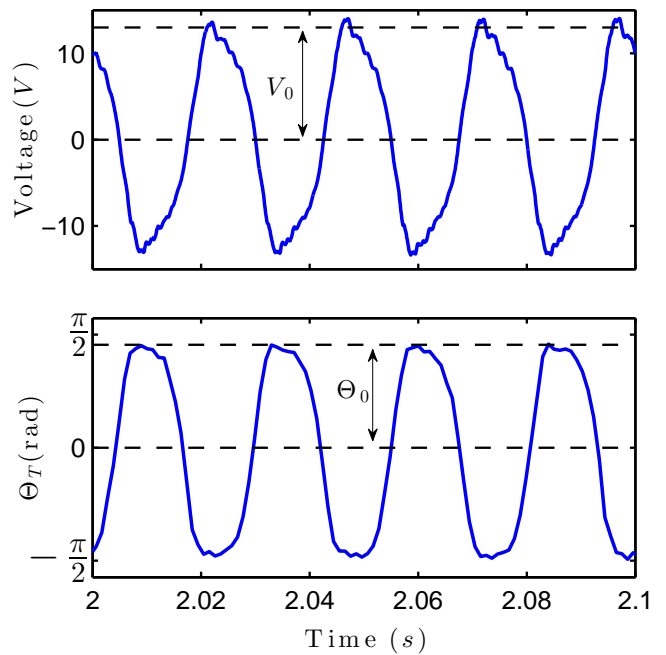}
\caption{\label{fig:volt_manip}Measurement of voltage $V$ and trailing edge angle $\Theta_T$ for $L=8$ cm and $U_\infty=12.3$ m/s.}
\end{figure}

The voltage is recorded using the DAQ (Figure~\ref{fig:volt_manip}), and $V_0$ is obtained by averaging the signal's peak values over the duration of the recording ($>$5s).
The flapping frequency $f$ is equal to the voltage signal's frequency, which is obtained using the Fourier transform of the recorded signal. In order to measure the trailing edge orientation $\Theta_T$, the flag's motion is recorded using a high-speed camera (Phantom\textsuperscript{\textregistered} v9) at 960 frames per second. In each frame, $\Theta_T$ is measured using ImageJ\textsuperscript{\textregistered} (Fig.~\ref{fig:measurement_Theta}) and $\Theta_0$ is obtained from these measurements (Fig.~\ref{fig:volt_manip}$b$).
Using this procedure, we obtained $B=2.03\times 10^{-5}$ N$\cdot$m$^2$ and $\chi=1.45\times 10^{-7}$ C. As a consequence, for the PVDF flag used here, $\alpha\sim 0.085$ for $L=6$ cm, and $\alpha\sim 0.1$ for $L=8$ cm. These values are retained for the numerical study.

\begin{figure}
\centering
\includegraphics{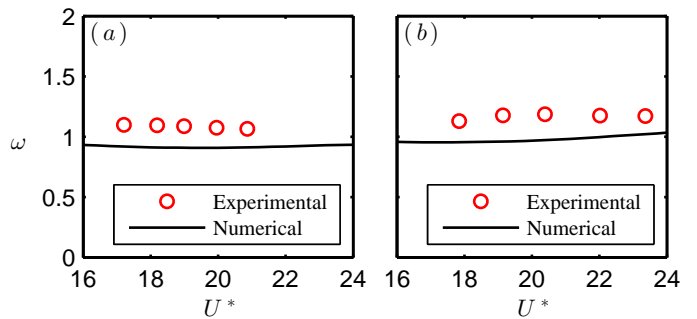}
\caption{\label{fig_Ustar_Freq_L006_L008_2exp_simu}Comparison of dimensionless flapping frequency $\omega$ as a function of dimensionless velocity $U^*$ obtained experimentally and numerically with $\alpha=0$. In $(a)$ Case A, $(b)$ Case B (see Table~\ref{tab:two_lengths}).}
\end{figure}

A first comparison between experiments and simulations is conducted for the case without piezoelectric coupling ($\alpha=0$). Figure~\ref{fig_Ustar_Freq_L006_L008_2exp_simu} shows the dimensionless flapping frequency $\omega$ as a function of dimensionless velocity $U^*$, obtained both experimentally and numerically, for $L=6$ cm and $8$ cm. Both results show a good agreement in terms of the flapping frequency, suggesting that the numerical model is capable of capturing the flag's essential dynamics. The fact that the numerical simulation slightly underpredicts the frequency compared with the experiments (Fig.~\ref{fig_Ustar_Freq_L006_L008_2exp_simu}) is likely due to the wind tunnel's weak but existing transverse confinement.\cite{belanger1995time}

\begin{figure}
\centering
\includegraphics{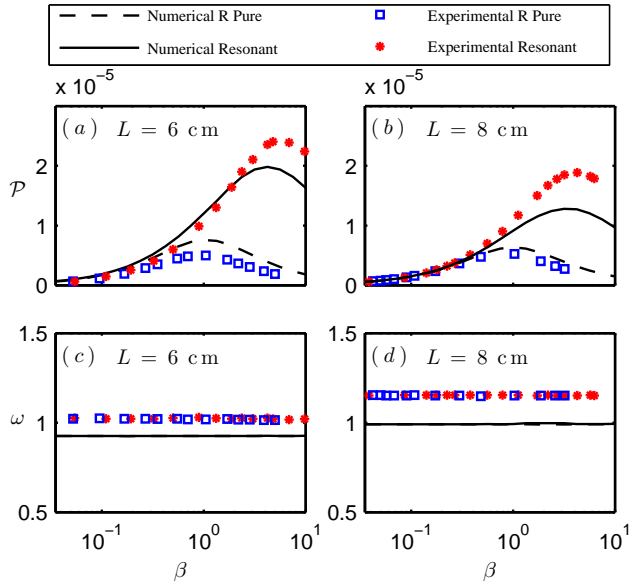}
\caption{\label{fig_Q_PVDF_exp_simu_L006_L008}$(a,b)$ Dimensionless harvested power $\mathcal{P}$ and $(c,d)$ flapping frequency $\omega$ as function of $\beta$ obtained both from experiments and nonlinear numerical simulations for cases A and B (see Table~\ref{tab:two_lengths}).}
\end{figure}

The energy harvesting performance is assessed both experimentally and numerically for two different effective lengths. The parameters corresponding to the two cases A and B are shown in Table~\ref{tab:two_lengths}.
\begin{table*}
\begin{tabular}{| c| c| c| c| c| c| c| c| c| c|}
\hline
  &$L$ (cm) & $U_\infty$ (m/s) & $f$ (Hz) & $\mathcal{L}$ (H) & $\alpha$ & $M^*$ & $H^*$ & $U^*$ & $\omega_0$\\\hline
Case A &6 	     & 20.9 			& 56.8 	   & 530 			   & 0.085	  & 0.410 & 0.417 & 17.91 & 1.06\\\hline
Case B &8 	     & 17.8 			& 41.0 	   & 1000 			   & 0.1	  & 0.547 & 0.313 & 21.18 & 1.14\\\hline
\end{tabular}
\caption{\label{tab:two_lengths}Parameter values of two piezoelectric flags and corresponding numerical simulation}
\end{table*}
Figure~\ref{fig_Q_PVDF_exp_simu_L006_L008} shows the dimensionless harvested power $\mathcal{P}$ (Fig.~\ref{fig_Q_PVDF_exp_simu_L006_L008}$a,b$) and  flapping frequency $\omega$ (Fig.~\ref{fig_Q_PVDF_exp_simu_L006_L008}$c,d$) as functions of the harvesting resistance $\beta$, obtained numerically and experimentally.
We observe that for both values of $L$, when the circuit is in resonance with the flapping flag, the harvested power increases considerably compared with the purely resistive case for almost all values of $\beta$. For a same resistance, the resonant circuit is able to harvest twice as much energy as the optimal purely resistive circuit. Meanwhile, the optimal resistance is larger when a resonant circuit is used, showing that resonant circuits maintain a satisfactory energy-harvesting performance even for large resistances for which little energy would be harvested with no inductance. At the optimal resistance for the resonant circuit, the system harvests 4 to 5 times more energy than the optimal resistive circuit. These results suggest that the presence of an inductance improves the energy harvesting performance by resonance. The fact that the circuit works in resonance produces a high voltage, leading to an enhanced harvested power. The same results are found using numerical simulations, showing a good agreement with the experiments in terms of the harvested power $\mathcal{P}$.

However, both experiments and numerical simulations suggest that resonance introduces little impact on the flags' flapping dynamics, as the flapping frequencies $\omega$ for both cases are almost identical for both types of circuit, and remain unchanged when varying $\beta$. The absence of a strong feedback induced by the inverse piezoelectric effect in the present work, as well as in other studies,\cite{demarquis:2011} is mainly due to the weak coupling of the chosen piezoelectric material.\cite{xia:2015} Another important factor is also the large internal resistance of the inductor ($\mathcal{R}_L\sim 1$~k$\Omega$), which introduces an additional damping to the system. This damping limits the voltage obtained at resonance,\cite{irwin2008basic} hence diminishes the harvested energy, and hinders the inverse piezoelectric effect. Reducing the inductor's internal resistance is therefore an important step toward a stronger feedback and a greater efficiency of the system.\cite{xia:2015}

Nevertheless, the present results confirm the validity of the nonlinear fluid-solid-electric model for predicting the energy-harvesting performance of the piezoelectric flag connected with a resistive-inductive circuit. This model can therefore be applied to more extensive parametric studies in order to identify potential piezoelectric feedback  effects and their influence on the energy harvesting, as performed in previous work.\cite{doare:2011, michelin:2013, xia:2015}

\bigskip
The experimental study presented in this paper confirms the benefit of a resonant circuit to the energy harvesting performance of the piezoelectric flag. Because the flag's flapping frequency is relatively low compared to a typical electric resonance frequency of a circuit with such a small capacitance, large inductances must be used but we show here that the required inductance is still within the commercially-available range, showing optimistic perspectives of the technology on the electrical aspect.

The results presented in this paper identify again the resonance as a key mechanism that improves the performance of energy-harvesting piezoelectric flags.\cite{xia:2015} This role of the resonance could also be extended to other kinds of vibration energy harvesters, whose performance can be improved by coupling them with resonant oscillators. Moreover, the potential feedback from the resonant oscillator to the vibration source would also be an effect acting in favor of the energy harvesting, and constitutes an interesting subject for further investigations.

Another perspective towards a better energy-harvesting performance lies in the choice of piezoelectric material. Composite materials, offering better piezoelectric coupling than polymers, while retaining satisfactory mechanical properties (flexibility, resilience) are promising candidates for future applications. Meanwhile, optimizing the number and the positioning of piezoelectric patches on the flag will also improve significantly the system's performance.\cite{pineirua2015influence}

This work was supported by the French National Research Agency ANR (Grant ANR-2012-JS09-0017).

\bibliographystyle{apsrev.bst}

\end{document}